\documentclass[aps,prl,twocolumn,groupedaddress,showpacs]{revtex4}

\usepackage{graphicx}

\begin{document}

\title{Phase Sensitive Recombination of Two Bose-Einstein Condensates on an Atom Chip}

\author{G.-B. Jo, J.-H. Choi, C.A. Christensen, T.A. Pasquini, Y.-R. Lee,
  W. Ketterle, and D.E. Pritchard}

\homepage[URL: ]{http://cua.mit.edu/ketterle_group/}

\affiliation{MIT-Harvard Center for Ultracold Atoms, Research
Laboratory of Electronics, Department of Physics, Massachusetts
Institute of Technology, Cambridge, MA 02139, USA}

\date{\today}

\begin{abstract}
The recombination of two split Bose-Einstein condensates on an
atom chip is shown to result in heating which depends on the
relative phase of the two condensates. This heating reduces the
number of condensate atoms between 10 and 40$\%$ and provides a
robust way to read out the phase of an atom interferometer without
the need for ballistic expansion. The heating may be caused by the
dissipation of dark solitons created during the merging of the
condensates.

\end{abstract}


\maketitle

Most experiments in atom interferometry use freely propagating atom
clouds~\cite{GBK97,PCC99}. Alternative geometries are confined-atom
interferometers where atoms are guided or confined in trapping
potentials~\cite{BPA97}, often realized by using atom
chips~\cite{FZM07}. These geometries are promising in terms of
compactness and portability, and also offer the prospect of
extending interrogation times beyond the typical 0.5 s achievable in
the atomic fountains. Such interferometers can be used to study
atom-surface interactions~\cite{PCC05} and Josephson
phenomena~\cite{AGF05}.

Many discussions of confined atom interferometers, inspired by
optical fiber interferometers, propose a readout by merging the two
separated clouds~\cite{HRH01,SJZ02,NHE04}. These discussions usually
assume non-interacting atoms~\cite{HRH01,ACF02} and don't address
the deleterious effects of atomic interactions, including dephasing,
collisional shifts, and phase
diffusion~\cite{CDR97,LYQ96,WWG96,JWP97,LSC98,JWR98}.  A recent
study showed that the recombination process is much more sensitive
to atomic interactions than the splitting process since merging
clouds with the opposite phase involves excited modes of the
recombined potential and can lead to exponential growth of unstable
modes~\cite{SZI03}. To circumvent these problems, previous
realizations of confined atom interferometry used ballistic
expansion of the two spatially independent condensates, which
decreases the atomic density before
overlap~\cite{SSP04,SJP05,SHA05,JSW07} or worked at very low atom
densities and pushed the clouds into each other with photon
recoil~\cite{GDH06,SWP07}. While this avoids the deleterious effects
of atom-atom interactions during the recombination, it lacks the
inherent simplicity and robustness of in-trap recombination.
Furthermore, in trap recombination, combined with dispersive,
\textit{in situ}, imaging ~\cite{AMV96}, could make it possible to
recycle the condensate for the next measurement cycle after
resetting the temperature through evaporating cooling. The detection
optics for \textit{in situ} imaging may even be integrated onto the
atom chip~\cite{SBC06}. Moreover, a trapped sample at high optical
density can be read out with sub-shot noise precision  using
cavity-enhanced atom detection~\cite{TLV06}.

In this letter, we show that in-trap recombination leads indeed to
heating of the atomic cloud.  However, this heating is
phase-dependent and can be used as a robust and sensitive readout of
the atom interferometer. The resulting oscillations of the
condensate atom number are dramatic (typically $\sim$25$\%$
contrast), occur over a wide range of recombination rates, and
permit high signal to noise ratios since they simply require a
measurement of the total number of condensate atoms in the trap.

The implications of phase-sensitive recombination extend beyond atom
interferometry. Recombination with uncontrolled phase was used to
replenish a continuous BEC ~\cite{CSL02} or to create vortices
~\cite{SWN06}. An extreme case of the merge process, where two
condensates are suddenly connected, has been studied by optically
imprinting a dark soliton into a single trapped
condensate~\cite{BBD99,DSF00}. Here we use methods of atom
interferometry to prepare two condensates with well-defined relative
phase and study the merging process for variable recombination
times.

Two special cases of the merging process can be exactly described
(Fig. 1). Two non-interacting separated condensates with the same
phase should adiabatically evolve into the ground state of the
combined potential, whereas a $\pi$-relative phase should result in
the lowest lying anti-symmetric state with excitation energy $N
\hbar\omega$ where $N$ is the total number of atoms in a trap and
$\omega$ is the the transverse frequency of the trapping potential.
The other limiting case is a merging process where a thin membrane
separates two interacting condensates until the potentials are
merged, and then is suddenly removed. For the 0-relative phase, the
merged condensate is in its Thomas-Fermi ground state.  For a
$\pi$-relative phase, however, the merged condensate contains a dark
soliton. Although the wave function differs from the ground state
only in a thin layer, the total energy of this excited state is
proportional to $N \hbar\omega$, as the lowest anti-symmetric state
in the non-interacting case.~\cite{foot:darksoliton}

Our working assumption is that the phase-sensitive excitation of the
cloud decays quickly, on the order of $\sim$1~ms in our system, and
leads to an increase in temperature on the order of
$\hbar\omega/{k_B}\simeq 100$~nK  for the case of
$\bigtriangleup\phi=\pi$, and less for other values of
$\bigtriangleup\phi$, where $k_B$ is the Boltzmann constant. The
parameters of our experiment were intermediate between limiting
cases of suddenness or adiabaticity, and we found a window of
recombination times for the phase-sensitive readout to which none of
these descriptions apply.

Bose-Einstein condensates of $\sim4\times10^{5}$ $^{23}$Na atoms in
the $|F=1, m_{F}=-1\rangle$ state were transferred into a magnetic
trap generated by the trapping wire on an atom chip and an external
bias field~\cite{SJP05}. The cloud had a condensate fraction
$\simeq$90\% and the temperature was $\sim$1/2 of the BEC transition
temperature, well above 0.1 when axial phase fluctuations are
excited. Using adiabatic rf-induced splitting~\cite{ZGT01,SHA05}, a
double-well potential in the horizontal plane was formed. Typically,
the separation of the two wells was d~$\sim6 \mu$m, the height of
the trap barrier was $U \sim h \times 10$~kHz, and the chemical
potential of the condensates, measured from the trap bottom, was
$\mu \sim h \times 6$~kHz, where $h$ is Planck's constant. In the
experiment, the coherence time of two separated condensates was at
least $\sim$50~ms ~\cite{JSW07}. The recombination of two split
condensates was realized by reducing the rf frequency as described
in Fig. 1(a), which decreases the trap barrier height. The merging
occurred slowly compared to the time scale determined by the radial
trap frequency ($\sim$1~kHz) to minimize mechanical excitation.

To monitor the energy increase after recombination, we measured the
central atom density during ballistic expansion. Phase-sensitive
collective excitations, in addition to mechanical excitations from
the splitting and merging processes,  heat the cloud and lower the
condensate fraction and, therefore, reduce the central density.  In
the experiment, the split condensates were held in the double well
potential for varying hold times, merged into a single potential,
and released by turning off the trapping potential within 30~$\mu$s.
After 8~ms time-of-flight, we measured the number of atoms in a
fixed area which is comparable to the size of (expanded)
Thomas-Fermi radius [dotted box in Fig. 2(c)]. While the total atom
number was conserved, the number within the fixed area decreased,
indicating that the temperature had increased. The fractional loss
of condensate atoms was obtained as the ratio of atom number after
recombination to the atom number before splitting.

\begin{figure}
\begin{center}
\includegraphics{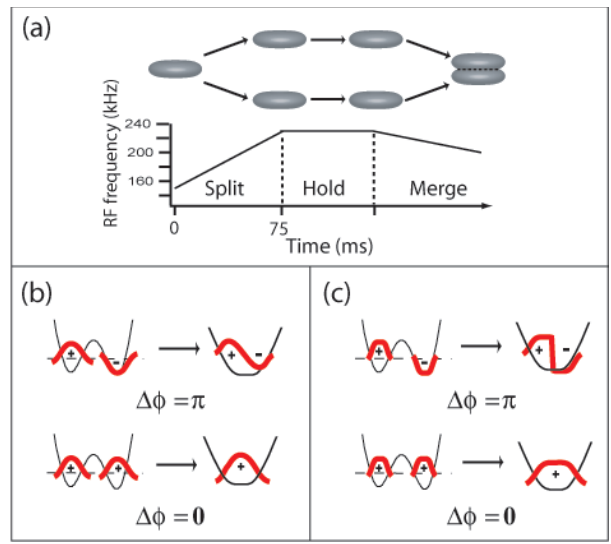}
\caption{(Color online) Schematic of the in-trap recombination
with a well defined relative phase.  (a) The phase-coherent
condensates were prepared using a radio frequency induced double
well potential on an atom chip~\cite{JSW07}. The splitting was
done within 75~ms by ramping up the rf frequency from 140~kHz to
225~kHz. During the hold time, the relative phase of two
independent condensates evolved with time at $\sim$500~Hz. After a
variable time, the double well potential was deformed into a
single well and the two trapped condensates were merged by
decreasing the rf frequency by 33~kHz over a variable
``recombination time''. The condensates started to spill over the
barrier after $\leq$10$\%$ of the recombination time or
$\sim$3~kHz decrease of the rf frequency. (b),(c) The merged
matter-wave functions are shown for the cases of an adiabatic
merger of non-interacting condensates and for a sudden merger of
interacting condensates.}
\end{center}
\end{figure}

\begin{figure}
\begin{center}
\includegraphics{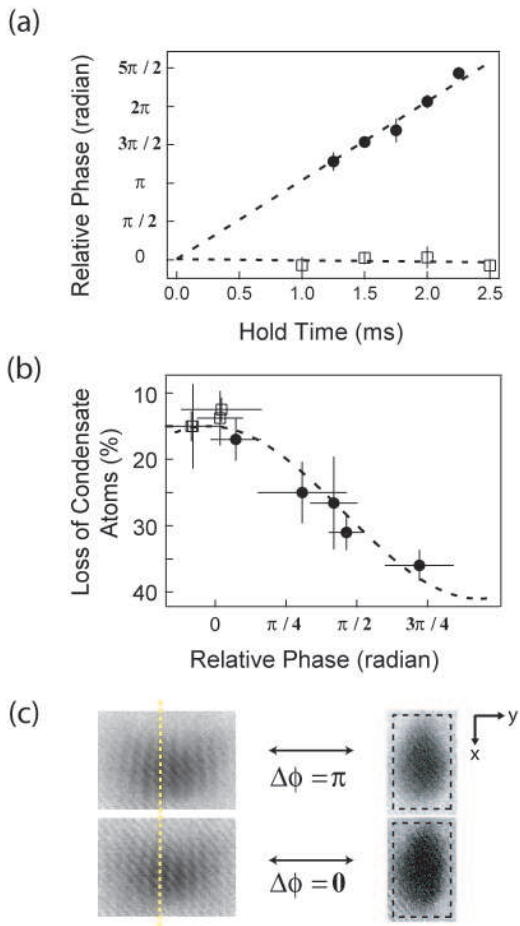}
\caption{(Color online) Phase-sensitive recombination of two
separate condensates. (a) The relative phase of two split
condensates was monitored for various hold time after splitting by
suddenly releasing the two condensates and observing interference
fringes. For the independent condensates (solid circle), the
evolution rate of the relative phase were determined from the linear
fit to be $\sim$500~Hz. For the weakly coupled condensates (open
square), the relative phase did not evolve. At 0~ms hold time, the
relative phase was set to zero for both cases. (b) For the same
range of delay times as in (a), the condensate atom loss after
in-trap recombination was determined. The relative phase (x-axis)
was obtained from interference patterns as in (a). The merging time
was 5~ms. (c) The matter-wave interference patterns (after 9~ms
time-of-flight) and absorption images of merged clouds (after 8~ms
time-of-flight) show the correlation between phase shift and
absorption signal. The field of view is 260 $\times$ 200 $\mu$m and
160 $\times$ 240 $\mu$m for matter-wave interferences and merged
clouds respectively.}
\end{center}
\end{figure}

\begin{figure}
\includegraphics{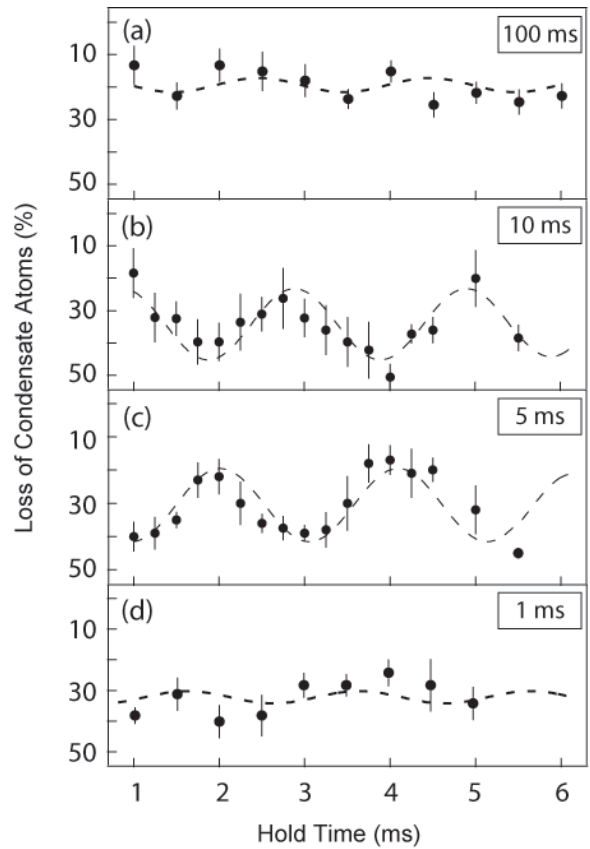}
\caption{Oscillations of condensate atom loss after recombination
reflecting the coherent phase evolution.   The condensate atom loss
was monitored during a variable hold time for the two split
condensates whose relative phase evolved at $\sim$500~Hz. The
merging was done for different values of the recombination time:
100~ms (a), 10~ms (b), 5~ms (c), and 1~ms (d). The dotted lines are
sinusoidal curves fitted with fixed frequency $\sim$500Hz. The
reproducible phase shift for the 5~ms and 10~ms data occurred during
the recombination process. The data points represent the average of
6 measurements.
 }
\end{figure}

\begin{figure}
\begin{center}
\includegraphics{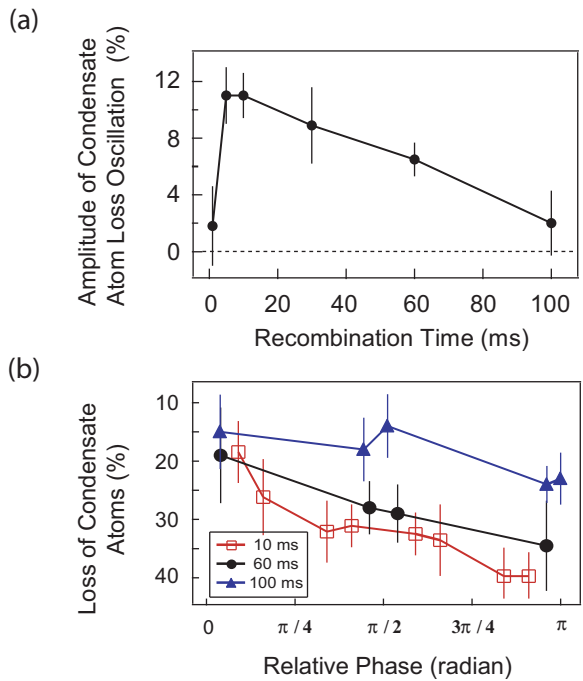}
\caption{ (Color online) Recombination time and atom loss. (a) The
amplitude of atom loss oscillations was determined for various
recombination times. (b) Assuming that minimum atom loss occurs at
0-relative phase of the  two condensates, relative phases were
obtained from the fitted atom loss oscillations in Fig. 3. }
\end{center}
\end{figure}

The fractional loss of condensate atoms was reproducible for a given
hold time, and observed to oscillate between 15$\%$ and 35$\%$ as a
function of hold time at a rate of 500~Hz (Fig. 2 and 3). The
observed oscillations are sinusoidal, although the non-linear
interactions can give rise to non-sinusoidal variations [15].
 To confirm
that this oscillatory heating was associated with the relative phase
of the split condensates, we measured the relative phase as the
spatial phase of the interference pattern when the split condensates
were suddenly released and interfered during ballistic expansion
[Fig. 2(a)]~\cite{SSP04}. The strong correlation between the two
measurements [Fig. 2(b)] is the central result of this paper.  As
the relative phase increased from $0$ to $\pi$, the atom loss after
recombination increased [Fig. 2(b)]; $\pi$-relative phase
($0$-relative phase) difference leads to maximum (minimum) loss of
condensate atoms.

The use of phase-sensitive recombination as a readout for an atom
interferometer is demonstrated in Fig. 3. The separated condensates
accumulate relative phase for an evolution time of up to 6~ms which
is read out after in-trap recombination. The phase-sensitive
recombination signal showed high contrast over a wide range of
recombination times [Figs. 3 and 4(a)]. The observed largest
amplitudes of condensate atom loss correspond to a change in
temperature on the order of $\sim$100~nK, in agreement with the
estimate in the introduction. This is testimony to the insensitivity
of the energy of phase-dependent excitations against changes in the
exact recombination parameters, and is promising for further
applications of chip-based atom interferometry.

The dependence of the condensate atom loss on the recombination time
allows us to speculate about different excitations caused by the
merging process.  The 1~ms recombination time shows little contrast
[Fig. 3(d)]. This time scale is comparable to the period of radial
oscillations, and one would expect breakdown of adiabaticity and
excitation of collective excitations independent of the relative
phase.  Significant loss ($\sim$30$\%$) was observed for all
relative phases and masked or suppressed any phase-sensitive signal.
The loss of contrast for the long recombination times could be
caused by relaxation of the phase-sensitive collective excitation
during the merging process when the condensates are connected only
by a region of low density, and soliton-like excitations have lower
energy.  An alternative explanation is the evolution of the relative
phase (at $\sim$500~Hz) during the effective recombination time.  In
a simple picture assuming a thin membrane being slowly pulled out
between the condensates, a phase evolution during this time would
create local solitons with phases varying between 0 and $\pi$.  This
could wash out the phase-sensitive signal to an average value. Since
the data for 100 ms recombination time show low loss [comparable to
the zero relative phase loss for faster recombination times, Fig.
3(d)], we favor the first explanation.  Furthermore, it is not clear
during what fraction of the ramp time of the rf frequency (called
the recombination time) the effective merging of the condensates and
the creation of a phase-sensitive collective excitation occurs. The
time between when the barrier equals the chemical potential and when
the the barrier reaches $\sim$70$\%$ of the chemical potential is
10$\%$ of the recombination time.  Another open question is what the
rate of phase evolution is at the moment of the merger. It is
plausible that during splitting, the condensates have the same
chemical potential, and that the observed difference is created only
when the condensates are further separated by ramping up the
barrier.  This would imply that during recombination, the situation
reverses, the chemical potential difference is reduced and reaches
near zero when the condensates merge. In any case, our work raises
intriguing questions for further experimental and theoretical
studies:  What kind of phase-sensitive excitations are created
during a merger process? How and when do they dissipate, and what
would happen when two condensates with different chemical potentials
are merged?

The present work demonstrates that interactions between atoms and
collective excitations are not necessarily deleterious to direct
recombination of separated trapped condensates that have acquired a
relative phase in atom interferometry. In contrast, the
phase-sensitive generation of collective excitations is used to
monitor the relative phase. This complements our previous work where
atomic interactions were shown to enhance the coherence time by
preparing a number squeezed state with the help of atomic
interactions during the beam splitting process~\cite{JSW07}. So the
merger between condensed matter and atomic physics goes both ways.
In recent years, atomic physics has developed powerful tools to
study many-body physics~\cite{AKK02}, and, as we have shown here,
many-body physics provides methods and tools to atom optics.

This work was funded by DARPA, NSF, and ONR. G.-B. Jo and Y.-R. Lee
acknowledge additional support from the Samsung foundation. We would
like to acknowledge Y. Shin for useful discussions, and M.
Vengalattore and M. Prentiss for atom chip fabrication.


\begin{thebibliography}{29}
\bibitem{foot:darksoliton}
In a layer on the order of the healing length $\xi$, the soliton has
a kinetic energy per atom proportional to $(1/\xi)^2$. With the
Thomas-Fermi radius proportional to $\hbar/(\omega \xi)$, the number
of atoms covered by the soliton is proportional to $N \omega \xi^2$,
and the total energy to $N \hbar\omega$.  If the condensates were
merged along their long axis, the soliton energy would be smaller by
the aspect ratio, that is, $\sim200$.
\end{thebibliography}
\end{document}